\documentclass[11pt,journal,draftcls,onecolumn,peerreviewca]{IEEEtran}
\usepackage{amsfonts}
\usepackage{amsmath}
\usepackage{amssymb}
\usepackage{bm}
\usepackage{xcolor,graphicx,float}
\usepackage{color, soul}
\usepackage{stfloats}
\usepackage[numbers,sort&compress]{natbib}
\usepackage[amsmath,thmmarks]{ntheorem}
\usepackage{subfigure}
\usepackage{theorem}
\usepackage{algorithm}
\usepackage{algorithmic}
\usepackage{graphicx}
\usepackage{subfigure}
\usepackage{pict2e}

\theoremheaderfont{\sc}\theorembodyfont{\upshape}
\theoremstyle{nonumberplain}
\theoremseparator{}
\theoremsymbol{\rule{1ex}{1ex}}

\begin{document}
\title{A comment on the ``A unified Bayesian inference framework for generalized linear models''}

\author{Jiang~Zhu \thanks{Jiang Zhu is with Ocean College, Zhejiang University (jiangzhu16@zju.edu.cn).}
}
\maketitle
\begin{abstract}
The recent work ``A unified Bayesian inference framework for generalized linear models'' \cite{meng1} shows that the GLM can be solved via iterating between the standard linear module (SLM) (running with standard Bayesian algorithm) and the minimum mean squared error (MMSE) module. The proposed framework utilizes expectation propagation and corresponds to the sum-product version \cite{Rangan1}. While in \cite{Rangan1}, a max-sum GAMP is also proposed. What is their intrinsic relationship? This comment aims to answer this.
\end{abstract}
{\bf Keywords:}
GLM, SLM, expectation propagation, MMSE, MAP
\newpage
\section{Max-sum GAMP}
According to \cite{Rangan1}, the output scalar estimation functions of sum-product GAMP (for MMSE estimation) are \cite{Rangan1}
\begin{align}\label{MMSEmeanout}
&g_{\rm out}(\hat{p},y,\tau_p)=(\hat{z}^0-\hat{p})/\tau_p,\\
&\hat{z}^0={\rm E}[z|\hat{p},y,\tau_p],\quad y\sim p(y|z),\quad z\sim {\mathcal N}(\hat{p},\tau_p).\notag
\end{align}
and
\begin{align}\label{MMSEout}
-g_{\rm out}^{'}(\hat{p},y,\tau_p)=(\tau_p-{\rm var}(z|\hat{p},y))/\tau_p^2.
\end{align}

While for the max-sum GAMP (for MAP estimation), the output scalar estimation functions are
\begin{align}\label{MAPmeanout}
g_{\rm out}(\hat{p},y,\tau_p)=(\hat{z}^0-\hat{p})/\tau_p,\\
\hat{z}^0=\underset{z}{\operatorname{argmax}}~F_{\rm out}(z,\hat{p},y,\tau_p).\notag
\end{align}
and
\begin{align}\label{MAPout}
-g_{\rm out}^{'}(\hat{p},y,\tau_p)=f_{\rm out}^{''}(\hat{z}^0,y)/(\tau_pf_{\rm out}^{''}(\hat{z}^0,y)-1),
\end{align}
where
\begin{align}
F_{\rm out}(z,\hat{p},y,\tau_p)\triangleq f_{\rm out}(z,y)-\frac{(z-\hat{p})^2}{2\tau_p},\quad f_{\rm out}(z,y)\triangleq \log p(y|z).
\end{align}

For MAP and MMSE, $\hat{z}^0$ is found via MAP or MMSE methods. Note that the output function (\ref{MAPmeanout}) of max-sum GAMP is basically the same as the output function (\ref{MMSEmeanout}) of sum-product GAMP. We now show that the output function (\ref{MAPout}) of max-sum GAMP can also be written in the form of sum-product GAMP (\ref{MMSEout}). To calculate $-g_{\rm out}^{'}(\hat{p},y,\tau_p)$ (\ref{MAPout}) for max-sum GAMP, we refer to the  sum-product GAMP methods.
By using Laplace approximation around $\hat{z}^0$ \cite{Mackay}, ${\rm var}_{\rm MAP}(z|\hat{p},y)$ is calculated as
\begin{align}\label{varMAP}
1/{\rm var}_{\rm MAP}(z|\hat{p},y)=-F_{\rm out}^{''}(\hat{z}^0,\hat{p},y,\tau_p)=-f_{\rm out}^{''}(\hat{z}^0,y)+1/\tau_p.
\end{align}
Substituting (\ref{varMAP}) in (\ref{MAPout}) and eliminating $f_{\rm out}^{''}(\hat{z}^0,y)$, we obtain
\begin{align}\label{MAPoutv2}
-g_{\rm out}^{'}(\hat{p},y,\tau_p)=(\tau_p-{\rm var}_{\rm MAP}(z|\hat{p},y))/\tau_p^2,
\end{align}
which has the same form as (\ref{MMSEout}). It has shown that the sum-product GAMP can be decomposed as SLM and MMSE module \cite{meng1}, as shown in Fig. \ref{SLM0313}. In the following, we show that max-sum GAMP can be decomposed as SLM and MAP module.
{\begin{figure}
  \centering
  \includegraphics[width=80mm]{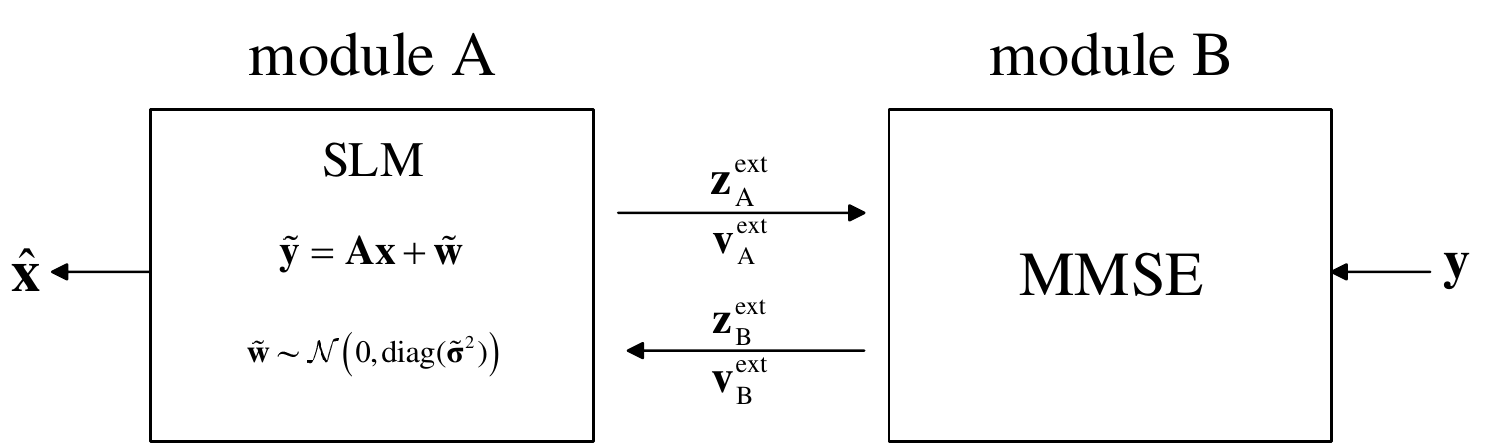}\\
\caption{A unified Bayesian inference framework proposed in \cite{meng1}. It is shown that utilizing the unified inference framework, many standard Bayesian inference algorithm can be extended to solve the GLM.}
\label{SLM0313}
\end{figure}}

It is shown in \cite{meng1} that $z_{\rm A}^{\rm ext}=\hat{p}$, $v_{\rm A}^{\rm ext}=\tau_p$, $z_{\rm B}^{\rm ext}=\tilde{y}$, $v_{\rm B}^{\rm ext}=\tilde{\sigma}^2$. For the AWGN channel, the output scalar estimation functions of GAMP \footnote{Sum-product GAMP and max-sum GAMP are the same in this setting.} are \cite{Rangan1}
\begin{align}\label{MMSEmeanout_AWGN}
g_{\rm out}(\hat{p},\tilde{y},\tau_p)=(\tilde{y}-\hat{p})/(\tilde{\sigma}^2+\tau_p),\\
\tilde{y}=z+{\mathcal N}(0,\tilde{\sigma}^2),\quad z\sim {\mathcal N}(\hat{p},\tau_p).\notag
\end{align}
and
\begin{align}\label{MMSEout_AWGN}
-g_{\rm out}^{'}(\hat{p},\tilde{y},\tau_p)=1/(\tilde{\sigma}^2+\tau_p).
\end{align}
According to expectation propagation (EP), $\tilde{y}$ and $\tilde{\sigma}^2$ is calculated as \cite{meng1}
\begin{subequations}\label{relat}
\begin{align}
\frac{1}{\tilde{\sigma}^2}+\frac{1}{\tau_p}=\frac{1}{{\rm var}_{\rm MAP}(z|\hat{p},y)},\\
\frac{\tilde{y}}{\tilde{\sigma}^2}+\frac{\hat{p}}{\tau_p}=\frac{\hat{z}^0}{{\rm var}_{\rm MAP}(z|\hat{p},y)}.
\end{align}
\end{subequations}
Substituting (\ref{relat}) in (\ref{MMSEmeanout_AWGN}) and (\ref{MMSEout_AWGN}) and eliminating ${\tilde{\sigma}^2}$ and $\tilde{y}$, one obtains (\ref{MAPmeanout}) and (\ref{MAPoutv2}). As a result, the sum-product GAMP can be decomposed as SLM and MAP module shown in \ref{SLM_MAP}.
{\begin{figure}
  \centering
  \includegraphics[width=80mm]{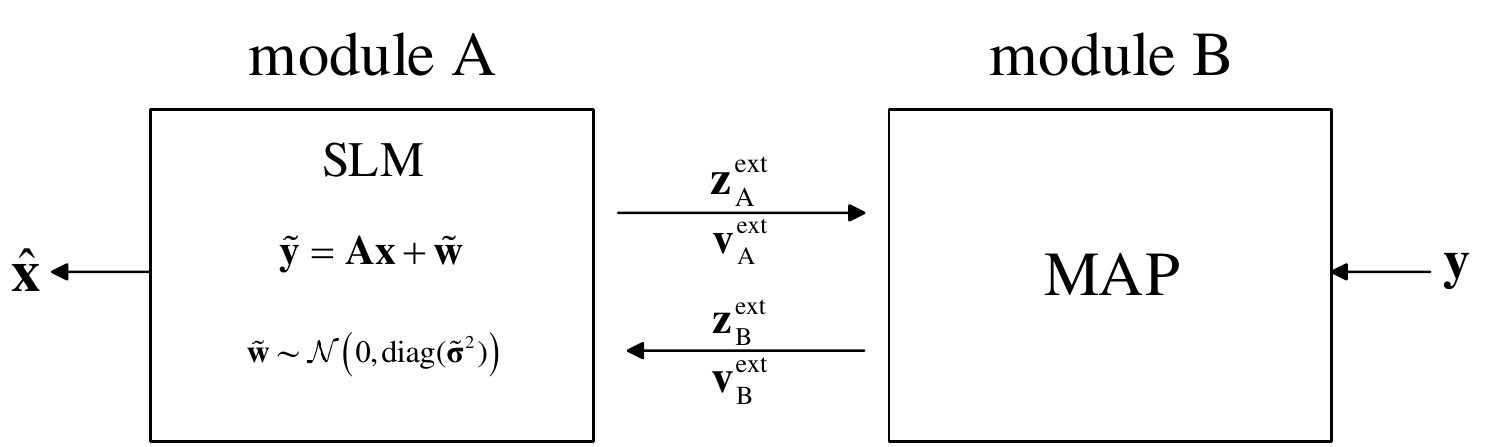}\\
\caption{A variant of the unified Bayesian inference framework \cite{meng1}. Here MAP is used instead of MMSE in module B.}
\label{SLM_MAP}
\end{figure}}
\section{Conclusion}
This note reveals the difference between max-sum GAMP and sum-product GAMP. Specifically, max-sum GAMP uses the MAP and Laplace approximation to calculate the MAP estimate and variance of $z$, while sum-product GAMP performs the MMSE and calculates the MMSE estimate and variance of $z$. For both max-sum GAMP and sum-product GAMP, EP is used to update the messages \cite{Minka, EPAMP}. 
\bibliographystyle{IEEEbib}
\bibliography{strings,refs}

\end{document}